# Topological Studies related to Molecular Systems formed soon after the Big Bang: HeH$_2^+$ as the Precursor for HeH$^+$

Narayanasami Sathyamurthy,[1] Michael Baer,[2] Satyam Ravi,[3] Soumya Mukherjee,[3] Bijit Mukherjee,[3] Satrajit Adhikari[3]

[1]) Jawaharlal Nehru Centre for Advanced Scientific Research, Bengaluru, India
[2]) The Fritz Haber Center for Molecular Dynamics, The Hebrew University of Jerusalem, Jerusalem, Israel.
[3]) School of Chemical Sciences, Indian Association for the Cultivation of Science, Kolkata, India.

# Abstract

In the early universe, following the nucleosynthesis, conditions were right for recombination processes to take place yielding neutral atoms H, He and Li. The understanding so far in astrophysics is that the first molecule to be formed was $HeH^+$ by radiative association (He + $H^+$ → $HeH^+$ + $h\nu$ and $He^+$ + H → $HeH^+$ + $h\nu$). The recent report by Güsten et al (Nature, **568**, 357, 2019) of detection of $HeH^+$ in planetary Nebula NGC 7027 confirms its presence, but it does not conclusively prove the origin of this species. To create molecules from free moving quasi-ions surrounded by an electronic cloud, the Born-Oppenheimer-Huang (BOH) theory furnishes two kinds of forces, namely, one that results from the Potential Energy Surfaces (PESs) and the other from Non-Adiabatic Coupling Terms (NACTs). Whereas the PESs are known to manage slow moving quasi-ions the NACTs, with their, frequently, infinitely large values at the vicinity of the singularities can control the fast moving quasi-ions. To achieve that the BOH equation indicates that the NACTs are affecting the fast moving quasi-ions directly and if they are attributed with **dissipative features** or in other words to behave as a **Friction Force** they indeed could serve (like any other ordinary friction) as moderators for the fast atomic/ionic species. It is proposed in the present paper that the triatomic $HeH_2^+$ was the precursor to $HeH^+$ and it could have been formed by the (He, H, H)$^+$ nuclei coming together under the electron cloud, facilitated by the NACTs between different electronic states acting as an astronomical friction force. This is possible because of the singularities in the NACTs for triatomic systems and NOT for diatomic systems. Although the existence of $HeH_2^+$ was established in the laboratory in 1996, it has not been detected in the interstellar media so far. But, there is no reason why it cannot be detected in near future.

# I. Introduction

Recently we started to study, topologically, molecular systems created shortly after the Big Bang (BB). [1-3] For this purpose we considered three molecular systems, namely, $H_3^{++}$, $H_3^+$ and $H_3$, with the emphasis to reveal to what extent $H_3^+$ − which according to our topological findings could be formed during the BB − **differs** from $H_3$ and $H_3^{++}$ which are not likely to be formed at all.

The way this study was done was by considering a plane (which in the present case is defined by the positions in configuration space (CS) of the three quasi-ions[4] that build-up the molecular system) and calculating the required Born-Oppenheimer-Huang (BOH) [5-7,8,9] NACTs (in this case the angular NACTs designated as $\tau_{\phi jj+1}(\phi|q,\mathbf{s})$). Here, **s** stands for the spatial coordinates of the three quasi-ions (see Fig. 1) and $(\phi, q)$ are polar coordinates to describe the distribution of the NACTs for a given situation in that plane. Since the tri-atomic NACTs are known to become singular at certain points − points which are associated with pairs of opposite Conical Intersections (*ci*s) − studies of this kind are associated with *ci*s.[8] The number of *ci*s is determined by various parameters that define the system, among them, the number of electronic states included in the study. Usually any two consecutive adiabatic states have at least one (common) singular point. Within the present study are considered NACTs associated with the three lowest states but sometimes, to achieve better converged results, NACTs due to four states and occasionally even five states had to be included. In most cases the *ci*s are located at the *equilateral* positions of the (tri-atomic) system and occasionally at the *isosceles* positions. Studies of that kind were reported recently.[1-3]

As mentioned earlier we are interested in the way molecules could be formed soon after the BB and this means situations where molecules are formed directly by free quasi-atoms wrapped by an electronic cloud.

In previous articles [1-3] we suggested that if we expect molecules to be formed soon after the BB the NACTs have to be associated with features that enable the creation of molecules directly from the free quasi-atoms in extreme *unfriendly* conditions, namely to form molecules from particles moving with enormous large velocities. This possibility was examined in detail with respect to the formation of $H_3^+$.

While discussing this issue in our previous articles we referred to the possibility that NACTs are blessed with features of the kind Gellman [10] and Zweig [11] attributed to their nuclear gluons. If that is the case, we could justify, at least heuristically the creation of tri-atomic molecules within the unfriendly atmosphere during the BB.

Here, we would like to refer to a different approach which could be more suitable for the situation we are facing here, namely, to attribute to the NACTs dissipative features or in other words to behave as a **Friction Force**.

The friction, in contrast to the ordinary forces that arise from the PESs, acts directly on the velocity of the moving body to slow it down. Friction forces are discussed in various text books on Classical Mechanics [12,13] and the typical case is a Hamiltonian of free mode vibrations affected by frictions proportional to the velocities. Consequently in Appendix I is *derived* a classical Hamiltonian for a single-linear-coordinate-system $x \{= (-\infty, \infty)\}$ affected by a force $kx$ and a friction force of the kind $\beta \dot{x}$, where $\dot{x}$ is the velocity:

$$\frac{d}{dt}\left[\frac{m}{2}\left(\dot{x}+\frac{1}{m}\beta x\right)^2+\frac{1}{2}kx^2\right]=\beta F(x,\dot{x},\ddot{x}) \qquad (1)$$

Here $F(x,\dot{x},\ddot{x})$ is given in the form

$$F(x,\dot{x},\ddot{x})=x\ddot{x}+\frac{d}{dt}\left(\frac{1}{2m}\beta x^2\right) \qquad (2)$$

As is noticed the main effect of the friction term is in disturbing the conservation of energy of the free Hamiltonian:

Recalling the BOH equation [8(a),3]

$$-\frac{\hbar^2}{2m}(\nabla+\mathbf{\tau})^2\psi+(\mathbf{u}-E)\psi=\mathbf{0} \qquad (3)$$

and comparing its structure with Eq.(1), it is noticed that indeed this equation suggests that the **τ** - matrix serves as a dissipative force (or a friction force). Eq. (3) does not include an explicit **source term** due to the friction so that the conservation of energy is maintained. As is well known the NACTs in Eq. (3) are formed by singularities. An important feature of friction is that during the slowing down process it produces energy as heat − in the present case the energy is produced as light or, better, as photons a feature that is believed to be detected, for instance during the production of HeH$^+$ (as will be discussed later).

Another fact to be noticed is that since the NACTs frequently become singular implies that the BOH friction may attain infinite large intensities and therefore could be rated among the **stronger** friction forces in the universe.

Based on what was said so far we present the following **summary:**
The BOH approach yields two forces which may affect the motion of the quasi-ions:
1. The PES which has its origin in the electrostatic interaction and therefore is a relatively weak force affecting the motion of these particles within a few eV range.
2. The NACTs which were formed after the BB could be identified as friction which affects straightforwardly the motion of the quasi ions. This process is enhanced once three or more quasi-ions are in a relative proximity because, in such a case, the NACTs attain, at certain regions in CS, infinite values. As a result, the fast moving quasi-ions are efficiently slowed down and in this way are forced to lose their excessive energy in a high rate so that the electrostatic forces may soon takeover.

Thus the conclusion is straightforward: If all that is needed for molecular ions to be formed after the BB is being equipped with the BOH NACTs then what is expected next is that the BOH approach produces the relevant NACTs at regions where they are needed most. A study like that was presented in Ref. 3 for $H_3^+$.

The fact that traces of the diatomic ion $HeH^+$ were recently detected in the Nebula NGC 7027 [14] brought us to extend our studies concerning the possibility that other tri-ionic molecules could have been formed after the BB. In the present article we intend to study the $HeH_2^+$ molecule which under certain conditions may serve as a precursor for the detected $HeH^+$.

Thus before completing this Section we briefly refer to diatomic molecules and to the possibility for them to be created, *straightforwardly*, after the BB. Since the diatomic NACTs are not singular, their ability to slow down the fast moving quasi–ions is limited, which essentially rules out the possibility for diatomic molecules to be created in this way (namely, straightforwardly). In other words their creation has to be an outcome of a chemical process. This possibility was mentioned while discussing the creation of $H_2$ and $H_2^+$ from $H_3^+$ and $H_3^{++}$.[1,2]

## II. The HeH$_2^+$ System as the Precursor for HeH$^+$

Thus, as before, we produce *circular* contours with their centers located at points along the equilateral line perpendicular to the axis formed by the two (hydrogenic) quasi-ions (see Fig. 1). Along these contours are calculated the angular components of corresponding NACTs [8(b)] as a function of the angle ϕ for a given contour (namely its radius and center). As before all these steps are undertaken with the aim of examining the possibility that the three **free** BOH quasi-ions, (He, H, H$^+$) are capable to form the HeH$_2^+$ molecular ion.

To the best of our knowledge such a molecule has not been detected in the interstellar media so far. However, the first spectral evidence for He…H$_2^+$ under laboratory conditions came from ion beam experiments in which mass selected ions were subjected to microwave radiation, in the presence of electric and magnetic fields.[15]

While a doublet was recorded around 21.8 GHz, with a spacing of 44.1MHz, a sextet was recorded around 15.2 GHz with a total spread of 13.2 MHz at low power and a second sextet with a spread of 16.8 MHz at a higher power. There were noticeable differences between *o*-H$_2$ and *p*-H$_2$ in the spectrum of the He…H$_2^+$ complex. There was a third resonance observed at 22.5 GHz and there were additional resonances observed in the range 123 - 168 GHz. Quantum chemistry studies indicate that these three quasi-ions are capable of forming a potential well as deep as − 0.37 eV supporting several bound states.[16-17] Time-dependent quantum mechanical wave packet studies have shown that there are several quasi-bound states as well.[18] The existence of the potential well is a *necessary* condition for poly-ionic molecules to be created − may be not always **indefinitely** but at least − for some short time.

As mentioned earlier, the key factors to be studied in this article are the NACTs. It is well known that deriving accurate and meaningful NACTs is a long and tedious task, which requires repeated assessments and tests. One of the more crucial tests for their relevance is the fulfillment of the molecular *quantization,* a feature frequently mentioned in our publications. [1-3, 8(c)] In general we distinguish between two-state, and multi-state results. In the present study we were forced to include up to four states to reach the required quantization (see also Ref. 17)

## III. Numerical Studies

### III.a Introductory comments

All ab initio calculations were performed by utilizing MOLPRO quantum chemistry package,[19] where state-averaged complete active space self-consistent field (SA-CASSCF) and multi-reference configuration interaction with singles and doubles(MR-CISD) methods were employed using the aug-cc-pVTZ (AVTZ) basis set including *s*, *p*, *d* functions of H and He. Calculations of PES are carried out by MR-CISD approach, whereas the NACTs have been computed by SA-CASSCF method. We have chosen complete active space (CAS) as (4o, 3e) i.e., three electrons distributed among four active orbitals. The lowest four doublet electronic states, namely, $1^2A_1$, $2^2A_1$, $3^2A_1$ and $4^2A_1$, and the NACTs between them were calculated using MR-CISD and CASSCF methodology, respectively.

### III.b Quantization as a Probe for Relevance of the NACTs

The numerical results are summarized in a Table given below (see Fig. 2). The various results are stored in **slots** where each set of three of them are arranged in a row that relates to a given situation. A situation is formed by the geometry of the three particles: the two hydrogen nuclei are kept fixed at points in CS: throughout this study the distance, $r$, between them is $r = 1.0$ Å (see Fig.1). The helium atom is applied as a test particle so that both the adiabatic potential energy curves − APECs − (calculated as a function of $R$, but for a fixed angle $\theta = \pi/2$ – see Fig. 1a) and the NACTS (calculated along circular contours defined in terms of a given radius $q$ and given centers in CS (see Fig. 1b)) are expressed by varying its position. Thus a situation is expressed in terms of $(r, \theta, q)$ and the center of the circle.

Next are discussed the slots: we have three types: A, B, C (see Fig. 2):

In each A-slot are given the APECs (all are identical) and the circular contours along which are calculated the NACTs. Thus in the A-slot of the first row is presented such a contour with a radius $q = 0.3$Å and its center at the (1,2) $ci$–point (thus, at $R = 0.54$Å); in the A-slot of the 2-nd row, is presented the contour with the same center but for a radius $q = 0.5$Å; in the A-slot of the 3-rd row is presented the contour with the center at $R = 1.0$ Å (thus a point in-between the (1,2) $ci$ and the (2,3) $ci$ points) for the (same) radius ($q = 0.5$ Å); in the A-slot of the 4-th row is presented a contour with the same center but for a radius $q = 0.7$ Å. In each B-slot of each row are presented the relevant NACTs, namely, $\tau_{\phi 12}(\phi|q,\mathbf{s}))$ and $\tau_{\phi 23}(\phi|q,\mathbf{s}))$ and in each C-slot the relevant ADT angles, thus $\gamma_{12}(\phi|q,\mathbf{s}))$ and $\gamma_{23}(\phi|q,\mathbf{s}))$ expected to be quantized. In the first two C-slots the contours surround only one $ci$, namely (1,2) $ci$ but in the two lower slots are presented contours that surround the two relevant $cis$, namely (1,2) $ci$ and (2,3) $ci$. In addition, in each of the C-slots is given the value of the corresponding topological phases $\alpha_{12}(q,\mathbf{s}))$ and, whenever available $\alpha_{23}(q,\mathbf{s})$ to emphasize, numerically, the extent these angles are indeed quantized (the topological phases are identified with the Berry Phase).[20-23] For a discussion on the Berry phases in the vicinity of conical intersections, the reader may see ref. 24.

**Summary:** It is important to emphasize the following: The better the ADT angles are quantized the more relevant are the NACTs. Indeed in all four cases presented here encouraging quantization was achieved – sometimes with three states and sometimes three state were not enough so we had to add the fourth state.

We remind the reader that our main interest lies in the (angular) NACTs which (based on what was claimed earlier) play the role of the astrophysical friction forces. If that is indeed the case the NACTs have to be intense enough to guarantee these frictions to be effective at **regions of the potential wells** (to enable this potential to form the tri-atom/ion molecule). From Fig. 3 it is well noticed, that these NACTs have to be intense along the collinear axis at the interval: $\Delta r \sim \{-2.0, 2.0 \text{ Å}\}$

## III.c A Detailed Study of the NACTs at the Potential Wells

This chapter is devoted to the intimate "engagement" between the potential wells responsible for the eventual creation of the $HeH_2^+$ molecule and the NACTs that are expected to form the conditions that this engagement becomes fruitful. In other words it is rather sensible to expect the NACTs to be intense at regions close enough to these wells in order to assist the electrostatic forces to take over.

Our study is done for the case the inter-atomic distance between the two hydrogen atoms is fixed – 1.0 Å. In this situation each of the two (symmetric) wells is located along the collinear axis ~1Å away from the corresponding hydrogen (see Fig. 3).

In Fig. 4a is presented the planar CS, the system of coordinates, the two fixed hydrogen atoms, the helium atom rotating along two the closed contours (with their center at the origin of the system of coordinates), the position of ten cis (five on each side the collinear axis) and the position of the two potential wells.

In Fig. 4b are presented two rows of slots. In each A-slots is given a circle (one with radius q=1Å and the other with radius q=1.5Å) surrounding the corresponding 4 *ci*s: (1,2)*ci*, (2,3)*c*i (3,4)*ci* and (4,5)*ci* as well as the five APECs.

The two B-slots contain the NACTs $\tau_{\phi12}(\phi|q)$ and $\tau_{\phi23}(\phi|q)$ and the two C-slots contain the NACTs $\tau_{\phi34}(\phi|q)$ and $\tau_{\phi45}(\phi|q)$ all as a function of $\phi$.

The conclusion of this part of our study is as follows: It is noticed that the NACTs are distributed uniformly throughout the (planar) CS and therefore are expected to be relevant in keeping the three quasi-ions in close proximity to each other and in this way support the potential well in, building-up the $HeH_2^+$ molecule.

## IV. Discussion and Conclusions

Nucleosynthesis seems to have occurred after 3 minutes after the initial expansion of the universe. Chemical synthesis[25] seems to have followed with recombination processes like

$He^{++} + e^- \rightarrow He^+ + h\nu$ (R1)

$He^+ + e^- \rightarrow He + h\nu$ (R2)

$H^+ + e^- \rightarrow H + h\nu$ (R3)

and the radiative association

$He + H^+ \rightarrow HeH^+ + h\nu$ (R4)

seems to have led to the formation of the first diatomic species.

An alternative channel of forming $HeH^+$

$He^+ + H \rightarrow HeH^+ + h\nu$ (R5)

seems to be several orders of magnitude less for red shift $z > 1000$.
The reaction

$He + H_2^+ \rightarrow HeH^+ + H$ (R6)

seems to be even less for $2000 < z < 200$. The reverse of reaction (R5)

$HeH^+ + h\nu \rightarrow He^+ + H$ (R7)

seems to have depleted $HeH^+$ for $z > 200$. In other words, reaction (R4) was a reversible process for $z > 200$.

The chemical reaction

HeH$^+$ + H → He + H$_2^+$ (R8)

seems to have played an important role in depleting HeH$^+$ for 200 < z < 10.

The dissociative recombination (DR)

HeH$^+$ + e → He + H (R9)

is believed to be a competing channel for reducing the population of HeH$^+$ in recent times (10 < z < 1). A quantitative comparison of the production and destruction rates of HeH$^+$ through different channels is given in Fig. 2 of ref. [26].

The detection by Güsten et al[14] of HeH$^+$ in the planetary nebula NGC 7027 through the recording of the spectrum of the J = 1→0 ground state rotational transition[27] provides the **long awaited confirmation** of the formation of HeH$^+$ in the early universe. Güsten et al[14] point out that the rate of radiative association in case of (R5) is ~ (1.4-6) x 10$^{-16}$ cm$^3$ s$^{-1}$ in the temperature range 5,000-20,000K and the rate of dissociative recombination in case of (R9) is ~ 3.0 x 10$^{-10}$ cm$^3$ s$^{-1}$ at 10,000K suggest much less HeH$^+$ than what is observed.

Novotny et al[28] conclude from experiments using cryogenic ion storage ring and merged electron beam that the DR rate used by Güsten et al is consistent with their observation.

Therefore, there is a need to identify additional channels of HeH$^+$ formation. Although 3-body collisions are generally less frequent than 2-body collisions, it is proposed that helium and hydrogen nuclei under the influence of electronic clouds could have formed [HeH$_2^+$]* in its multiple electronic states stabilized by NACTs acting as some kind of a frictional force. They, in turn, could emit photons and come to lower energy states and also dissociate to form HeH$^+$.[29]

Interestingly, the formation of H$_3^+$ in the early universe seems to have been accounted for[23] and we have proposed [1-3] that its stability arises from the NACTs. Zicler et al[30] had proposed the formation of HeH$_3^+$ through the reaction

He + $H_3^+$ → $HeH_3^+$ (R10).

The authors discounted the formation of $HeH_2^+$ + H because that channel is a few eV higher in energy.

In this publication we consider for the first time the possibility that the $HeH_2^+$ molecule was created shortly after the BB. This molecule, to the best of our knowledge has not been detected in the interstellar media so far, but that doesn't mean that it could not be formed for a short time and then dissociate via one, or more, of the four possible given processes:

$$HeH_2^+ \rightarrow \begin{cases} He^+ + H_2 & (A1) \\ He + H_2^+ & (A2) \\ HeH^+ + H & (A3) \\ He + (H+H)^+ & (A4) \end{cases}$$

According to this scheme we encounter creations of, eventually, three different diatomic molecules. However reaction (A1) is not likely to happen because of its exceptionally high endothermicity and out of the two other reactions the reaction (A2) seems more promising because of its larger exothermicity and the reaction (A3) is slowed down due to the barrier formed by the endothermicity. Still recalling that our required (final) product is $HeH^+$ (and not $H_2^+$), reaction route (A3) is probably the preferred route as $HeH^+$ is formed directly (although with more difficulties). This is said because forming $HeH^+$ via route (A2) requires one additional reaction (to form $HeH^+$), namely, (R6). However applying this reaction encounters two difficulties: not only the same barrier formed by the endothermicity as in reaction (A3) but also a reagent (in this case, He) which is still moving with high velocities.

Another chain of possible reactions to form HeH$^+$ is to consider what happens while creating H$_3^+$. In our recent article we showed that from a topological point of view, this molecule could be formed after the BB and having this molecule will eventually yield the following two competing reactions:

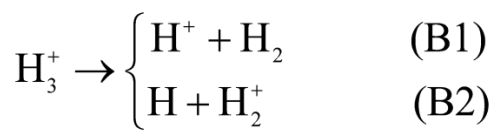

$$H_3^+ \to \begin{cases} H^+ + H_2 & (B1) \\ H + H_2^+ & (B2) \end{cases}$$

However to achieve the product HeH$^+$ we need again to activate reaction (R6) which was mentioned earlier creates fundamental difficulties.

**The conclusion of all that:**

Based on the BOH equation the chain of reactions to form the diatomic molecule HeH+ surrounded by the BB atmospheric conditions is as follows

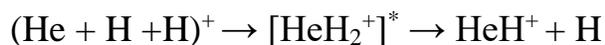

(He + H +H)$^+ \to$ [HeH$_2^+$]$^*$ $\to$ HeH$^+$ + H

**Acknowledgement:**

NS is an Honorary Professor at Jawaharlal Nehru Centre for Advanced Scientific Research, Bengaluru. Satyam Ravi acknowledges the Indian Association for Cultivation of Science, Kolkata for research fellowship. Soumya Mukherjee (File No. SPM-07/080(0250)/2016-EMR-I) and Bijit Mukherjee (File No. 09/080(0960)/2014-EMR-I) thank the Council of Scientific and Industrial Research, New Delhi for research fellowships. Satrajit Adhikari is thankful to Science and Engineering Research Board, New Delhi for research funding through Project No. EMR/2015/001314.

# Appendix I

The basic one dimensional (classical) Equation describing a body moving under the influence of force $kx$ and affected by a friction force $\beta \dot{x}$ is given in the form [12]

$$m\ddot{x} + \beta \dot{x} = -kx \qquad (I.1)$$

Multiplying the equation by $\dot{x}$ we get the expression:

$$m\ddot{x}\dot{x} + \beta \dot{x}^2 = -kx\dot{x} \qquad (I.2)$$

Next introducing explicitly the derivative with respect to time we find for the l.h.s.:

$$m\dot{x}\ddot{x} + \beta \dot{x}^2 = \frac{m}{2}\frac{d}{dt}(\dot{x}^2) + \beta \frac{d}{dt}(x\dot{x}) - \beta x\ddot{x} \qquad (I.3)$$

or,

$$m\dot{x}\ddot{x} + \beta \dot{x}^2 = \frac{d}{dt}\left(\frac{m}{2}\dot{x}^2 + \beta x\dot{x}\right) - \beta x\ddot{x} \qquad (I.4)$$

In the same way r.h.s. of Eq. (I.2) is found to become:

$$kx\dot{x} = \frac{1}{2}k\frac{d}{dt}(x^2) \qquad (I.5)$$

Substituting Eqs. (I.4) and (I.5) in Eq. (I.2) yields (following minor changes):

$$\frac{d}{dt}\left(\frac{m}{2}\dot{x}^2 + \beta x\dot{x}\right) + \frac{1}{2}k\frac{d}{dt}(x^2) = \beta x\ddot{x} \qquad (I.6)$$

Forming the squaring:

$$\frac{m}{2}\frac{d}{dt}\left(\dot{x}+\frac{1}{m}\beta x\right)^2 + \frac{1}{2}k\frac{d}{dt}\left(x^2\right) = \beta x\ddot{x} + \frac{d}{dt}\left(\frac{1}{2m}\beta^2 x^2\right) \qquad (I.7)$$

or:

$$\frac{d}{dt}\left[\frac{m}{2}\left(\dot{x}+\frac{1}{m}\beta x\right)^2 + \frac{1}{2}kx^2\right] = \beta F(x,\dot{x},\ddot{x}) \qquad (I.8)$$

where

$$F(x,\dot{x},\ddot{x}) = x\ddot{x} + \frac{d}{dt}\left(\frac{1}{2m}\beta x^2\right)$$

When the friction coefficient $\beta$ becomes zero the conservation of energy is restored.

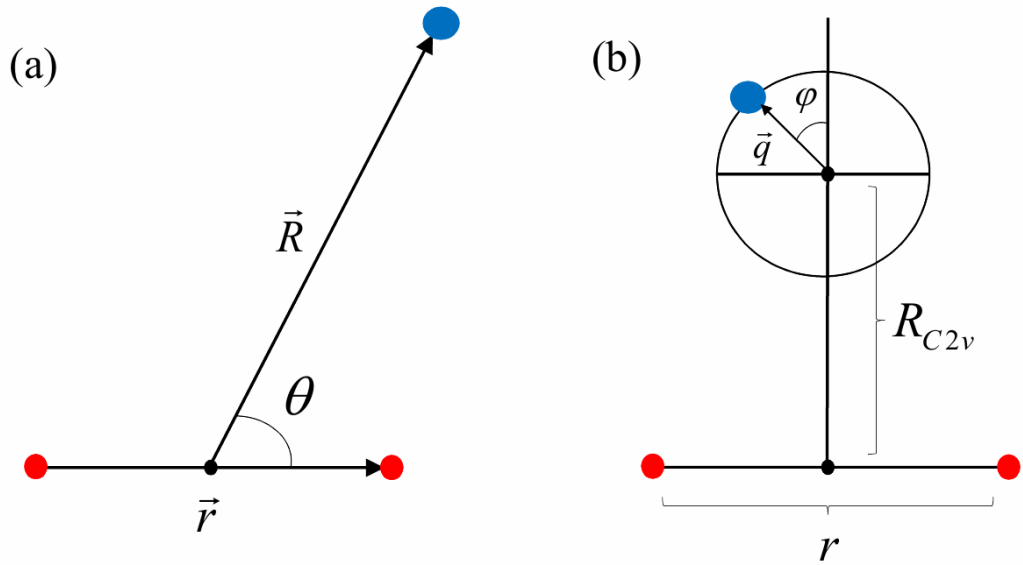

**Figure 1:** Schematic picture of the system of coordinates and various points of location: (a) The Jacobi system of coordinates, namely, $r$, $R$ and $\theta$; (b) The point of (1,2) or (2,3) *ci* is at $R = R_{C2v}$ and $\theta = \pi/2$, which is the center of the circle with the radius $q$.

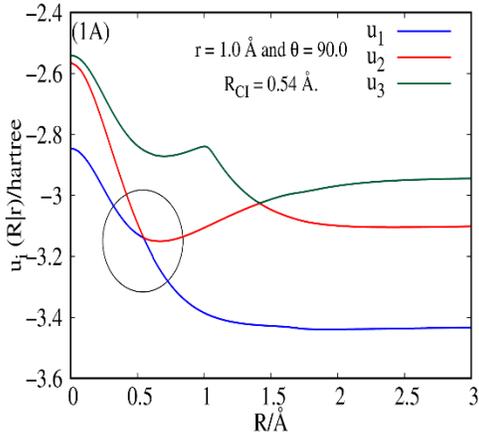
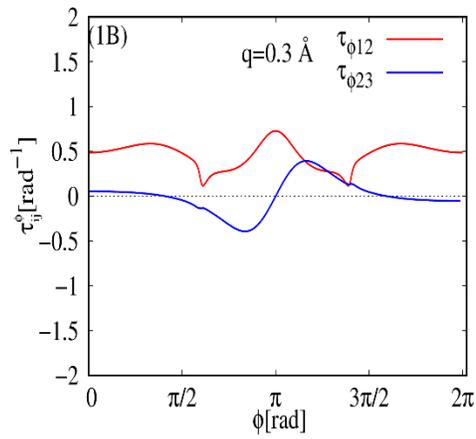
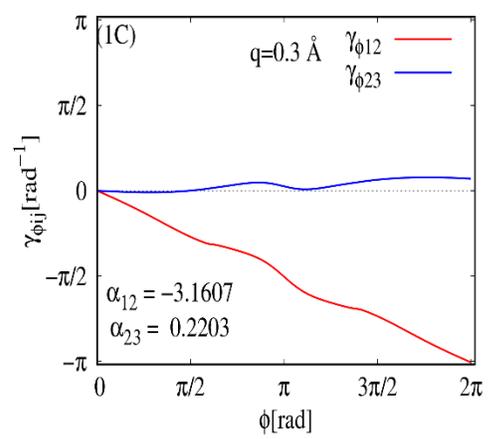
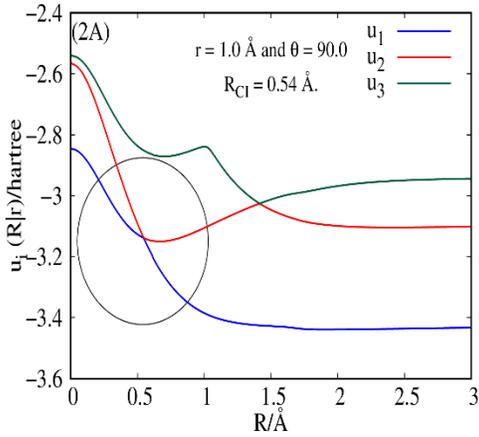
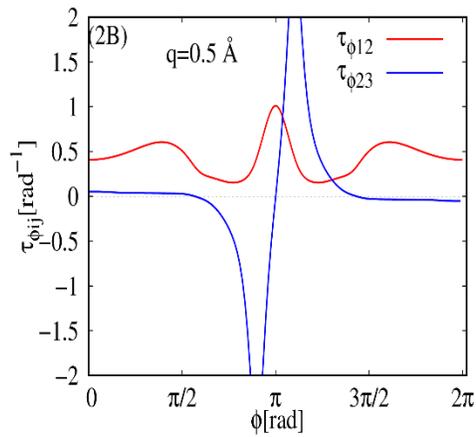
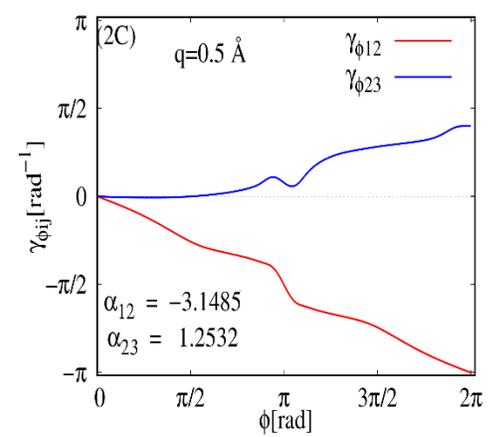
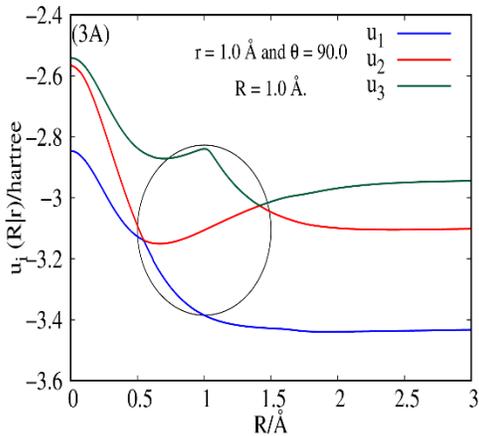
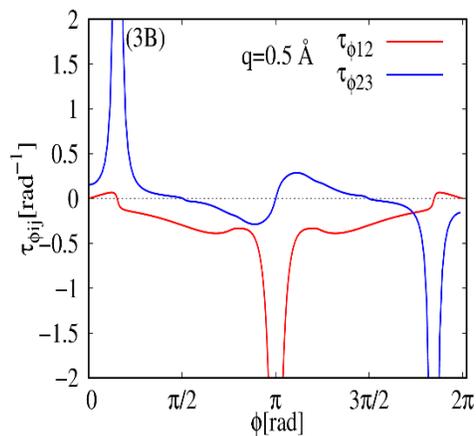
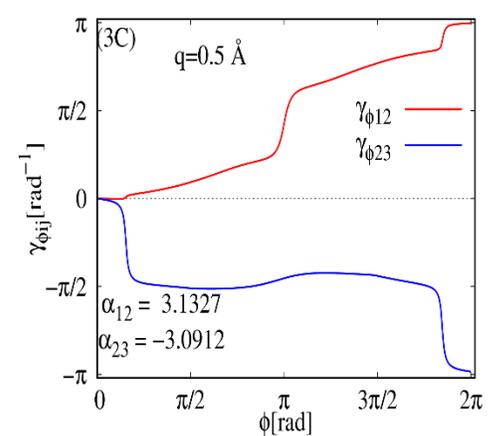
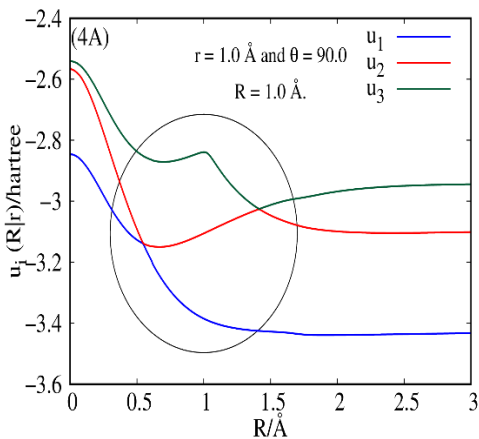
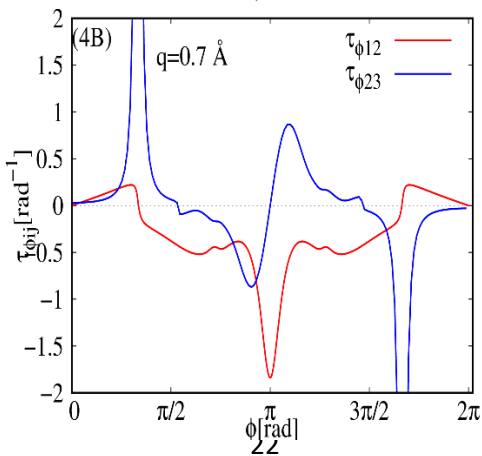
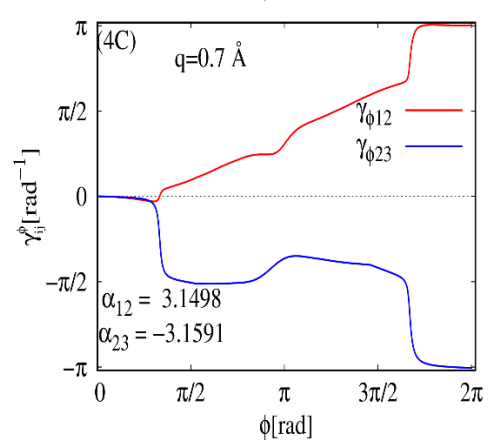

**Figure 2:** Panel 1: (A-C) show results for $HeH_2^+$ calculated at $r = 1.0$ Å, $R = 0.54$ Å for radius $q = 0.3$ Å; Panel 2: (A-C) show results for $HeH_2^+$ calculated at $r = 1.0$ Å, $R = 0.54$ Å for radius $q = 0.5$ Å; Panel 3: (A-C) results for $HeH_2^+$ calculated at $r = 1.0$ Å, $R = 1.0$ Å for radius $q = 0.5$ Å; Panel 4: (A-C) results for $HeH_2^+$ calculated at $r = 1.0$ Å, $R = 1.0$ Å for radius $q = 0.7$ Å; The panels (1A, 2A, 3A, 4A) represent three adiabatic potential energy curves (APECs), $u_i(R|r)$, $i = 1,3$, as a function of $R$ for fixed $r = 1.0$ Å. The panels (1B, 2B, 3B, 4B) display the two relevant NACTs $\tau_{\phi 12}(\phi|q,\mathbf{s})$ and $\tau_{\phi 23}(\phi|q,\mathbf{s})$ as a function of $\phi$, calculated for circles with centers located at $R = 0.54$ Å and $R = 1.0$ Å. The panels (1C, 2C, 3C, 4C) depict the two relevant ADT angles $\gamma_{12}(\phi|q,\mathbf{s})$ and $\gamma_{23}(\phi|q,\mathbf{s})$ as a function of $\phi$ and the relevant (printed) topological phases $\alpha_{ij}$.

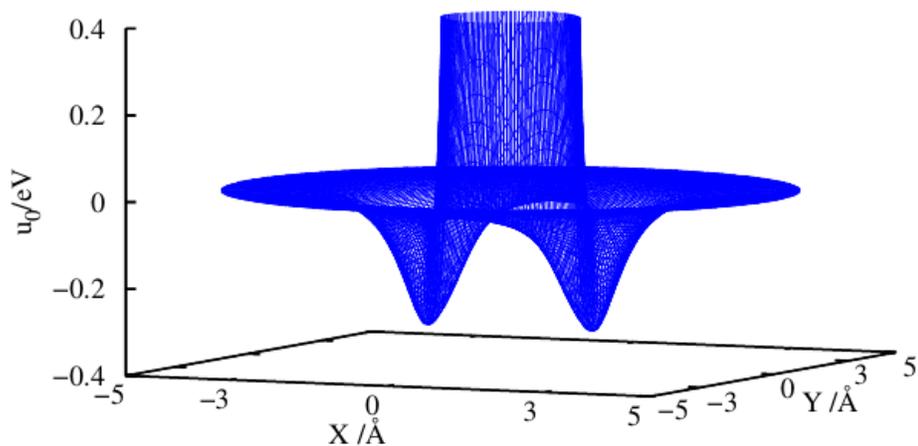

**Figure 3:** The adiabatic potential energy surfaces of the HeH$_2^+$ system over *X-Y* plane, where X = $R\cos(\theta)$ and Y = $R\sin(\theta)$ for $r = 1.0$ Å. The positions of the minima for the formation of HeH$_2^+$ are X = 1.5, Y = 0 and X = -1.5, Y = 0 with well-depth - 0.34 eV, where the energy is scaled with respect to the asymptote of well separated He and H$_2^+$.

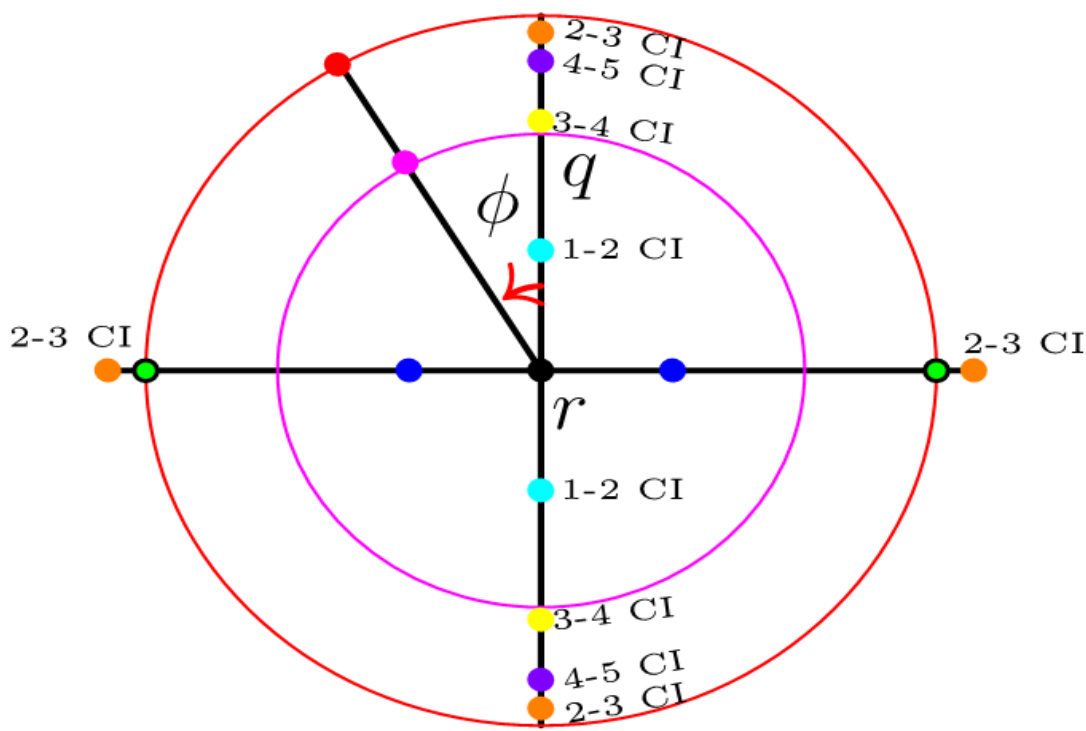

**Figure 4a:** The contour generation for $HeH_2^+$ system with the center at $r = 1.0$ Å, $R = 0.0$ Å for $q = 1.0$ (magenta colour) and 1.5 Å (red colour). Blue dots indicate the position of two hydrogens, whereas He atom is shown by magenta colour and red colour in two different contours. The approximate position of minima of $HeH_2^+$ is shown by green dots at $r = 1.0$ Å, $X = 1.5$ Å, $Y = 0.0$ Å and $X = -1.5$ Å, $Y = 0$ Å, where $X = R\cos(\theta)$, $Y = R\sin(\theta)$ with $\theta = 0$ and $\pi$. Total ten *ci* points are shown in this figure. The position of two symmetric $C_{2V}$ (1,2) *ci*s is indicated by cyan colour at $r = 1.0$ Å $X = 0.0$ Å, $Y = 0.54$ Å and $X = 0.0$ Å, $Y = -0.54$ Å. The positions of two symmetric $C_{2V}$ (2,3) *ci*s are indicated by orange colour at $r = 1.0$ Å, $X = 0.0$ Å, $Y = 1.45$ Å and $X = 0.0$ Å, $Y = -1.45$ Å. Another two symmetric collinear (2,3) *ci*s are indicated by orange colour at $r = 1.0$ Å, $X = 1.6$ Å, $Y = 0.0$ Å and $X = -1.60$ Å, $Y = 0.0$ Å. The positions of two symmetric $C_{2V}$ (3,4) *ci*s are indicated by yellow colour at $r = 1.0$ Å, $X = 0.0$ Å, $Y = 1.01$ Å and $X = 0.0$ Å, $Y = -1.01$ Å. The positions of two symmetric $C_{2V}$ (4,5) *ci*s are indicated by violet colour at $r = 1.0$ Å, $X = 0.0$ Å, $Y = 1.42$ Å and $X = 0.0$ Å, $Y = -1.42$ Å.

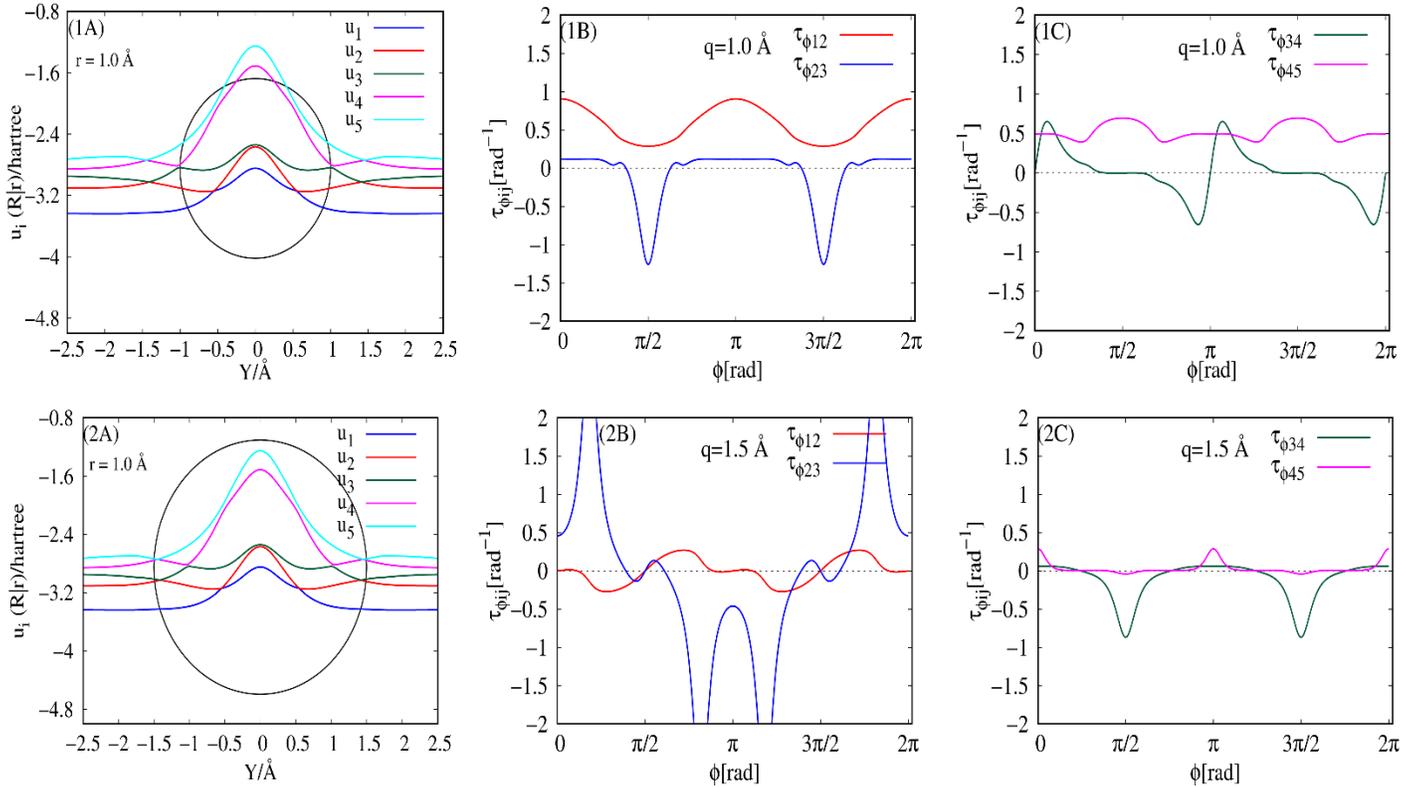

**Figure 4b:** Panel 1: (A-C) show results for $HeH_2^+$ calculated at $r =1.0$ Å, $R= 0.0$ Å for radius $q = 1.0$ Å; Panel 2: (A-C) show results for $HeH_2^+$ calculated at $r =1.0$ Å, $R= 0.0$ Å for radius $q = 1.5$ Å; The panels (1A, 2A) represent five adiabatic potential energy curves (APECs), $u_i(R|r)$, $i =1,5$, as a function of Y coordinate ($Y = R\sin(\theta)$, where $\theta = \pi/2$ and $3\pi/2$) for fixed $r =1.0$ Å. The panels (1B, 2B) display the two relevant NACTs $\boldsymbol{\tau}_{\phi12}(\phi|q,\mathbf{s})$ and $\boldsymbol{\tau}_{\phi23}(\phi|q,\mathbf{s})$ as a function of $\phi$, calculated for circles with centers located at $R = 0.0$ Å with $q = 1.0$ Å and 1.5 Å. The panels (1C, 2C) depict the other two relevant NACTs $\boldsymbol{\tau}_{\phi34}(\phi|q,\mathbf{s})$ and $\boldsymbol{\tau}_{\phi45}(\phi|q,\mathbf{s})$ as a function of $\phi$ calculated for circles with centers located at $R = 0.0$ Å with $q = 1.0$ Å and 1.5 Å.